\documentclass[preprint,floatfix,superscriptaddress]{revtex4}
\usepackage{graphicx}
\usepackage{bm}
\usepackage{amsmath}

\newcommand{\be}{\begin{equation}}
\newcommand{\ee}{\end{equation}}

\begin{document}
\title{Current-induced domain wall motion in a nanowire with perpendicular magnetic anisotropy}
\author{Soon-Wook Jung}
\affiliation{PCTP and Department of Physics, Pohang University of
Science and Technology, Pohang, Kyungbuk 790-784, Korea}
\author{Woojin Kim}
\affiliation{Department of Materials Science and Engineering, Korea Advanced Institute of Science and Technology,
Daejeon 305-701, Korea}
\author{Taek-Dong Lee}
\affiliation{Department of Materials Science and Engineering, Korea Advanced Institute of Science and Technology,
Daejeon 305-701, Korea}
\author{Kyung-Jin Lee}
\altaffiliation[Corresponding e-mail: ]{kj\_lee@korea.ac.kr}
\affiliation{Department of Materials Science and Engineering, Korea
University, Seoul 136-701, Korea}
\author{Hyun-Woo Lee}
\altaffiliation[Corresponding e-mail: ]{hwl@postech.ac.kr}
\affiliation{PCTP and Department of Physics, Pohang University of Science and Technology, Pohang, Kyungbuk 790-784, Korea}
\date{\today}

\begin{abstract}
We study theoretically the current-induced magnetic domain wall motion
in a metallic nanowire with perpendicular magnetic anisotropy.
The anisotropy can reduce the critical current density of the domain wall motion.
We explain the reduction mechanism and
identify the maximal reduction conditions.
This result facilitates both fundamental studies and device applications
of the current-induced domain wall motion.
\end{abstract}

\maketitle

Spin-polarized electrical currents in ferromagnets can transfer
their spin angular momentum to local magnetizations via the
$s$-$d$ exchange interaction and generate
torques~\cite{Berger,Slonczewski} on local magnetizations. This
spin transfer torque (STT) received considerable attention in view
of both fundamental physics
research~\cite{Grollier,Klaui1,Tatara1,Zhang,Thiaville} and
applications~\cite{Parkin,Allwood}.

In a ferromagnetic nanowire, the STT can generate motion of
magnetic domain walls (DWs). For conventional metallic
ferromagnetic nanowires, which have the in-plane magnetic
anisotropy (IMA), experiments~\cite{Yamaguchi,Klaui2,Hayashi}
found such current-induced DW motion when the current density $J$
in the nanowire is larger than a certain threshold value $J_{\rm
c}$ of the order of $10^8$ A/cm$^2$. This value is too high; At
such current densities, the Joule heating generates considerable
thermal fluctuations~\cite{Yamaguchi2,Laufenberg,You}, making
fundamental studies of the STT difficult. Furthermore device
applications~\cite{Sze} require $J_{\rm c} < 10^7$ A/cm$^2$ at
room temperature. Thus both for fundamental studies and device
applications, it is crucial to reduce $J_{\rm c}$.

Recently there are experimental~\cite{Ravelosona} and
theoretical~\cite{Fukami} indications that $J_{\rm c}$ may be
considerably lower in a metallic nanowire with the perpendicular
magnetic anisotropy (PMA). However it remains unclear how the PMA
can lower $J_{\rm c}$. We aim to answer this question in this
Letter.

We consider a nanowire with the wire width $w$ along the $y$-axis
and thickness $t$ along the $z$-axis ($w>t$). We use the
Landau-Lifshitz-Gilbert (LLG) equation with the STT term,
\begin{equation}
\begin{split}
\dot{\mathbf{m}}&=-\gamma \mathbf{m} \times \mathbf{H}_{\rm eff}
+\alpha \mathbf{m} \times \dot{\mathbf{m}} +b_J
(\hat{\mathbf{J}}\cdot \nabla) \mathbf{m}\\
& \quad -c_J \mathbf{m} \times (\hat{\mathbf{J}}\cdot
\nabla)\mathbf{m}, \label{llg}
\end{split}
\end{equation}
where $\mathbf{m}$ is the unit vector of the local magnetization,
$\gamma$ is the gyromagnetic ratio, $\alpha$ is the Gilbert
damping parameter, $\hat{\mathbf{J}}$ is the unit vector of the
local current density, and $\mathbf{H}_{\rm eff}$ is the effective
magnetic field. $b_J=P \mu_B J/e M_s$ is the magnitude of the
adiabatic STT~\cite{Tatara1}, where $e$ is the electron charge,
\textit{P} is the spin-polarization of the ferromagnet, $\mu_B$ is
the Bohr magneton, and $M_{\rm s}$ is the saturation
magnetization. $c_J$ is the magnitude of the nonadiabatic
STT~\cite{Zhang,Thiaville} with the non-adiabaticity represented
by the dimensionless parameter $\beta \equiv c_J/b_J$. $\beta$ is
independent of $J$ and estimated~\cite{Zhang} to be of the order
of $10^{-2}$.

To get an insight into the main physics of the PMA, we first
develop an analytical model based on a one-dimensional (1D)
approximation. Its results will be later verified by performing
the micromagnetic simulations of the LLG equation
[Eq.~(\ref{llg})], which are known to provide a reliable
description of nanoscale magnetization dynamics~\cite{Lee,Seo}.

For a ferromagnet with the PMA we have
\begin{equation}
\mathbf{H}^{\rm eff}=\frac{2A}{M_{\rm s}} \frac{\partial^2
\mathbf{m}}{\partial x^2}+\frac{2K_{U} m_z}{M_{\rm s}}
\mathbf{e}_z + \mathbf{H}_{\rm dipole}, \label{heff}
\end{equation}
where $A$ is the exchange stiffness constant and $K_U$ is the PMA
constant that allows the easy axis (along the $z$-axis) to be
perpendicular to the wire-plane ($x$-$y$ plane). To describe the
demagnetization effects, we consider the magnetostatic
dipole-dipole interaction field given by $\mathbf{H}_{\rm
dipole}(\mathbf{r})=M_{\rm s} \int d^3r' N(\mathbf{r}-\mathbf{r'})
\mathbf{m(r')}$, where the components of the matrix $N$ are given
by $N_{xx}(\mathbf{r})=-[1-3x^2/|\mathbf{r}|^2]/|\mathbf{r}|^3$,
$N_{xy}(\mathbf{r})=3xy/|\mathbf{r}|^5$. Other components of $N$
are defined in a similar way.

We also assume that the DW maintains the following shape during
the DW motion; $m_z(\mathbf{r},t)=\tanh[(x-q)/\lambda]$,
$m_x(\mathbf{r},t)= \cos \psi {\rm sech}[(x-q)/\lambda]$,
$m_y(\mathbf{r},t)= \sin \psi {\rm sech}[(x-q)/\lambda]$, where
$\lambda$ is the equilibrium DW width obtained from 1D
micromagnetic simulations. In this rigid DW motion
approximation~\cite{Jung} the DW dynamics is described by two
dynamical variables, the DW position $q(t)$ and the DW tilting
angle $\psi(t)$.

By using the procedure developed by Thiele~\cite{Thiele}, one can
then derive, from the LLG equation, the equations of motion for
the two collective coordinates $q$ and $\psi$,
\begin{align}
\lambda \dot{\psi}-\alpha \dot{q}&=c_J-(\gamma \lambda / 2M_{\rm
s})f_{\rm pin}, \label{eomcc1}\\
\dot{q}+\alpha \lambda \dot{\psi}&=-b_J - (\gamma \lambda / M_{\rm
s}) K_{\rm d} \sin2\psi, \label{eomcc2}
\end{align}
where the pinning force $f_{\rm pin}$ is related to the DW energy
per unit cross-sectional area $u_{\rm tot}$ ($f_{\rm
pin}=-\partial u_{\rm tot}/\partial q$) representing pinning
potential in the presence of extrinsic defects in a nanowire. Here
$K_{\rm d}$ is the effective wall anisotropy given by
\begin{equation}\label{Kd}
K_{\rm d}=K_y-K_x,
\end{equation}
where $K_i= -\frac{M_{\rm s}^2}{4S\lambda} \iint d^3r d^3r'
N_{ii}(\mathbf{r}-\mathbf{r}') {\rm sech}\frac{x}{\lambda}{\rm
sech}\frac{x'}{\lambda}$ ($i=x,y,z$) and $S$ is the
cross-sectional area. $K_{\rm d}$ represents the magnetostatic
energy difference between two types of transverse DWs, the Bloch
DW ($\mathbf{m}\parallel \mathbf{e}_y$ at the DW center) and the
Neel DW ($\mathbf{m} \parallel \mathbf{e}_x$ at the DW center).

\begin{figure}
\includegraphics[width=1.0\columnwidth]{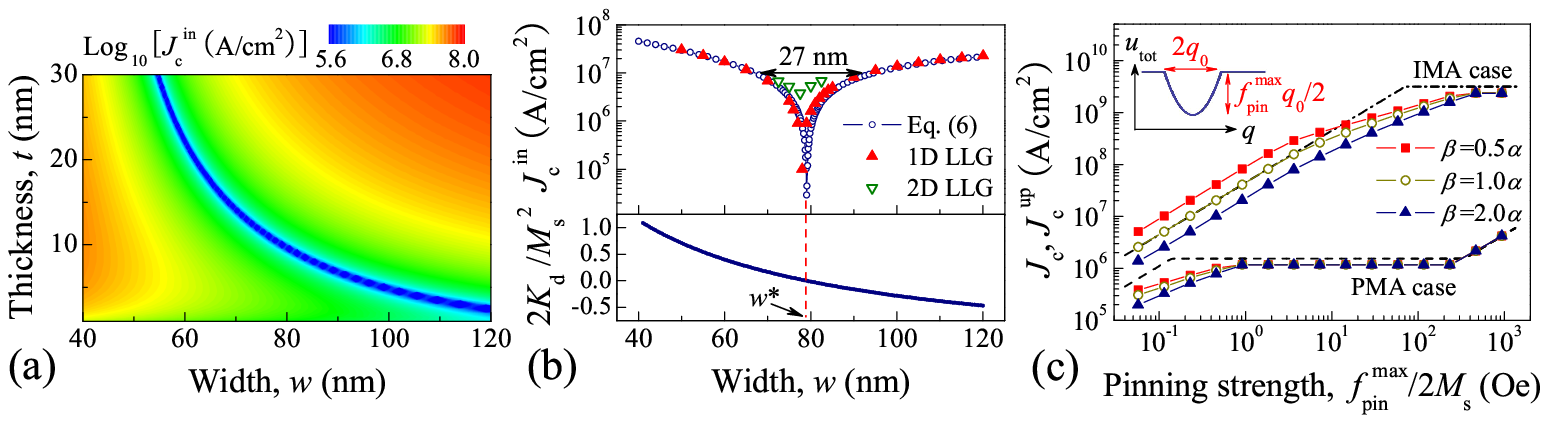}
\caption{(Color online) (a) $J_{\rm c}^{\rm in}$ from
Eq.~(\ref{jcin}) as a function of $w$ and $t$. (b) Upper panel:
$J_{\rm c}^{\rm in}$ defined in Eq.~(\ref{jcin}) vs $J_{\rm c}$
obtained from micromagnetic simulations of the 1D and 2D LLG
equation. Lower panel: $K_{\rm d}/M_{\rm s}^2$ as a function of
$w$. $t=10$ nm in both panels. (c) The dashed (dash-dotted) line
shows $J_{\rm c}^{\rm up}$ [Eq.~(\ref{jcup})] for a PMA (IMA)
nanowire with $\beta=\alpha$ and $J_{\rm c}^{\rm in}=1.6\times
10^6 (3.15\times 10^9)$ A/cm$^2$. Symbols shows $J_{\rm c}$
obtained from numerical simulations of Eqs.~(\ref{eomcc1}) and
(\ref{eomcc2}). The result is in reasonable agreement with $J_{\rm
c}^{\rm up}$ for $\beta=\alpha$. Inset: Spatial profile of $u_{\rm
tot}$ with $q_0=3\lambda$. The following parameters are used:
$\alpha=0.02$, and $P=0.7$, $A=1.3\times 10^{-6}$ erg/cm, $K_U=1.5
\times 10^6 (0 \times 10^6)$ erg/cm$^3$, and $M_{\rm s}=400 (800)$
emu/cm$^3$ for PMA (IMA) nanowire.} \label{analytical}
\end{figure}

Before we demonstrate its implications for a general case, we
first consider a defect-free nanowire ($f_{\rm pin}=0$) with
$c_J=0$. $J_{\rm c}$ in this case is given~\cite{Tatara1} by
$J_{\rm c}^{\rm in}$
\begin{equation}\label{jcin}
J_{\rm c}^{\rm in} \equiv \frac{e \gamma \lambda}{P \mu_{\rm B}}
|K_{\rm d}|.
\end{equation}
Figure~\ref{analytical}(a) shows $w$ and $t$ dependence of $J_{\rm
c}^{\rm in}$. Note that $J_{\rm c}$ falls below $10^7 {\rm
A/cm^2}$ in a wide range of $w$ and $t$. Since all material
parameters used in Fig.~\ref{analytical} are similar to those for
permalloy except for the PMA constant $K_U$, this reduction in
$J_{\rm c}^{\rm in}$ should be attributed to the PMA. To check the
validity of this prediction, we also perform micromagnetic
simulations of the LLG equation and excellent agreement is found
[Fig.~\ref{analytical}(b) upper panel].

This reduction in $J_{\rm c}^{\rm in}$ becomes especially
effective when $w$ is tuned to a $t$-dependent special value
$w^*(t)$, at which $K_{\rm d}$ reverses its sign
[Fig.~\ref{analytical}(b) lower panel] and near $w^*(t)$, $J_{\rm
c}^{\rm in}$ $(\propto |K_{\rm d}|)$ is strongly suppressed. The
sign reversal of $K_{\rm d}$ implies that $w^*$ is the equilibrium
phase boundary between the Bloch DW and Neel DW. For transverse
DWs in an IMA nanowire, on the other hand, $K_{\rm d}$ is given by
$K_z-K_y$ and since $K_z$ is always larger than $K_y$, $K_{\rm d}$
in the IMA case is always positive in a conventional nanowire
geometry with $w>t$. This difference between a PMA nanowire and an
IMA nanowire illustrates a crucial role played by the PMA.

Next we consider a general case with $f_{\rm pin}\neq 0$ and $c_J
\neq 0$. After some calculation, one can obtain an upper bound
$J_{\rm c}^{\rm up}$ of $J_{\rm c}$~\cite{Tatara2},
\begin{equation}\label{jcup}
J_{\rm c}^{\rm up}\equiv {\rm max}\left[{\rm min}(J_{\rm c}^{\rm
in},J_{\rm c}^{\rm ex}),\alpha\beta J_{\rm c}^{\rm ex}\right],
\end{equation}
where $J_{\rm c}^{\rm ex}\equiv (\gamma \lambda e/2P\mu_{\rm
B})(f_{\rm pin}^{\rm max}/\beta)$ and $f_{\rm pin}^{\rm max}$
represents the maximum value of $f_{\rm pin}$. The dashed line in
Fig.~\ref{analytical}(c) shows $J_{\rm c}^{\rm up}$ as a function
of $f_{\rm pin}^{\rm max}$ for a PMA nanowire. For the case
$J_{\rm c}^{\rm in}=1.6 \times 10^6 {\rm A/cm^2}$,
Fig.~\ref{analytical}(c) also shows $J_{\rm c}$ determined from
numerical simulations of Eqs.~(\ref{eomcc1}) and (\ref{eomcc2})
with the pinning potential energy $u_{\rm tot}$ modelled by a
finite ranged harmonic potential [Fig.~\ref{analytical}(c) inset].
A few remarks are in order. Firstly, both $J_{\rm c}$ and $J_{\rm
c}^{\rm up}$ exhibit plateaus near $J_{\rm c}^{\rm in}$ in a wide
range of $f_{\rm pin}^{\rm max}$. Secondly, $J_{\rm c}$ depends on
$\beta$ only in the weak pinning regime ($f_{\rm pin}^{\rm
max}/2M_{\rm s}< 1$ Oe) and the $\beta$ dependence essentially
disappears in the intermediate (plateau) and strong (above
plateau) pinning regimes. This behavior is consistent with the
prediction of Eq.~(\ref{jcup}). Thirdly, a recent
experiment~\cite{Tanigawa} with a PMA nanowire finds the depinning
magnetic field of about 500 Oe for a field-driven DW motion. When
this value is used as an estimation of $f_{\rm pin}^{\rm
max}/2M_{\rm s}$, one finds $J_{\rm c}^{\rm up} \sim J_{\rm
c}^{\rm in} \sim 10^6$ A/cm$^2$. Thus Fig.~\ref{analytical}(c)
demonstrates that the reduction of $J_{\rm c}^{\rm in}$ via the
PMA indeed leads to the reduction of $J_{\rm c}$. As a comparison,
results for an IMA nanowire are also given in
Fig.~\ref{analytical}(c). Differences from the PMA case are
evident.

\begin{figure}
\includegraphics[width=0.7\columnwidth]{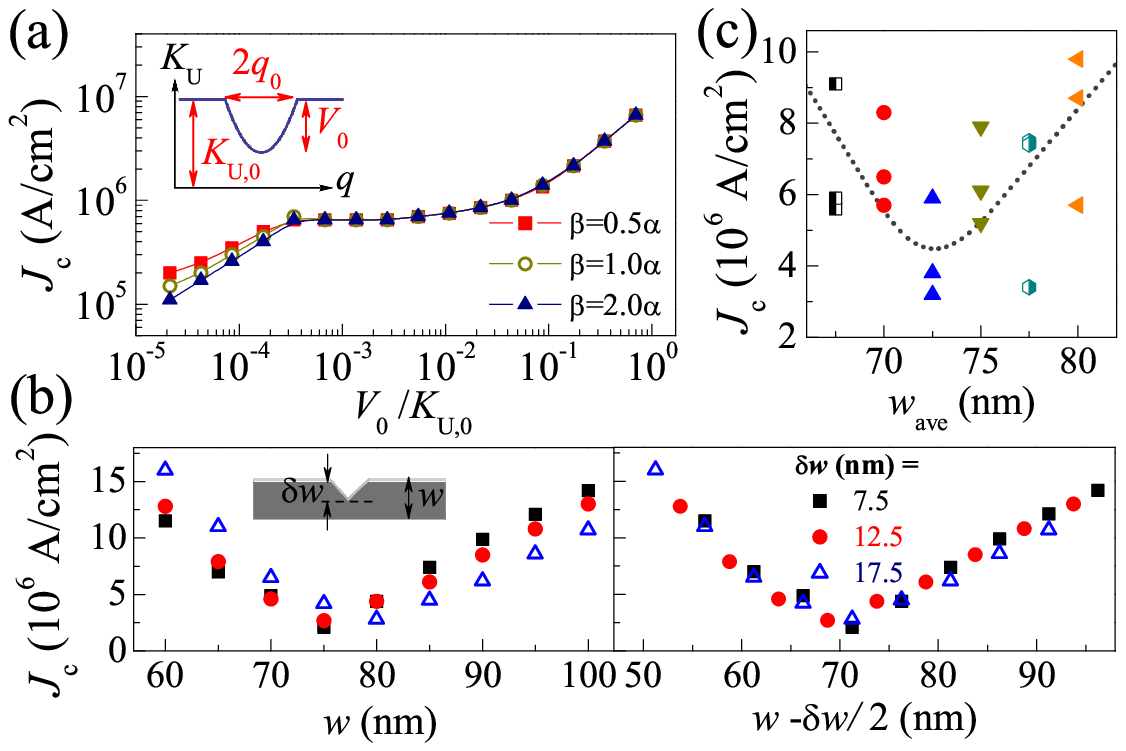}
\caption{(Color online) Micromagnetic simulation results of the
LLG equation. (a) Effects of the magnitude fluctuation of the PMA
constant $K_U$ for $w=77$ nm and $t=10$ nm. Upper inset: Spatial
profile of $K_U$ with $q_0=37.5$ nm. (b) Effects of a notch.
Inset: Schematic of a notch. (c) Combined effects of edge
roughness (2.5 nm for each edge) and PMA fluctuations (Gaussian
magnitude fluctuations of 5\% and direction fluctuations of
5$^\circ$ for each cell of size 2.5 nm $\times$ 2.5 nm $\times$
$t$ nm). For each $w_{\rm ave}$, three realizations of the
randomness are considered (The dotted line is a guideline).
$\beta=\alpha$ and $t=10$ nm. All other parameters are the same as
those in Fig.~\ref{analytical}.} \label{simulation}
\end{figure}

Next we present micromagnetic simulation results of the LLG
equation. Various sources of $f_{\rm pin}$ are examined.
Figure~\ref{simulation}(a) shows $J_{\rm c}$ obtained from the 1D
LLG equation for a situation where the magnitude of the PMA
constant $K_U$ fluctuates from its bulk average value $K_{U,0}$
with the maximum deviation given by $V_0$. Note that the result is
remarkably similar to that in Fig.~\ref{analytical}(c).
In good quality PMA samples, $V_0/K_{U,0}$ is
reported~\cite{Thomson} to be less than 0.1, for which we obtain
$J_{\rm c} \approx 10^6$ A/cm$^2$. Figure~\ref{simulation}(b)
shows the effect of a notch investigated with the two-dimensional
(2D) LLG equation. In a wide range of $w$, $J_{\rm c}$ falls below
$10^7$ A/cm$^2$ despite the notch formation. Note that for $w \geq
80$ nm, $J_{\rm c}$ decreases as the notch depth  $\delta w$
increases. This strange behavior is not due to the locally
enhanced current density near the notch, since this effect should
be stronger for $w \leq 80$ nm. Instead it is due to the fact that
$J_{\rm c}^{\rm in}$ is determined by an \textit{effective wire
width} that a DW senses. When $J_{\rm c}$ is plotted as a function
of $w- \delta w/2$, an estimation of the effective width, this
strange behavior disappears and $J_{\rm c}$ is now almost
independent of $\delta w$, in agreement with the prediction
$J_{\rm c}^{\rm up}=J_{\rm c}^{\rm in}$ in the plateau range in
Fig.~\ref{analytical}(c). Figure~\ref{simulation}(c) shows $J_{\rm
c}$ for a PMA nanowire with the edge roughness and with the PMA
fluctuations. Although values of $J_{\rm c}$ are somewhat
scattered with the realizations of the randomness, $J_{\rm c}$
still remains below $10^7$ A/cm$^2$ in a wide range of the average
width $w_{\rm ave}$~\cite{Jc-minimum-comment}.

Here we remark that all demonstrations for the reduction of
$J_{\rm c}$ assume the proper tuning of $w$ and $t$ to achieve the
reduced $J_{\rm c}^{\rm in}$. A recent experiment on the PMA
nanowire~\cite{Tanigawa} found $J_{\rm c}=1.0 \times 10^8$
A/cm$^2$ without such tuning. We suggest that the tuning of $w$
and $t$ can reduce $J_{\rm c}$. Another
experiment~\cite{Ravelosona} found indications of the enhanced STT
efficiency in a PMA spin valve. However the measurement was still
restricted to the thermally assisted creep regime with extremely
low DW velocity (average $v_{\rm dw}<10^{-8}$ m/s). According to
our calculation (not shown), much higher velocity ($v_{\rm dw}
\sim 10$ m/s) can be achieved at $J \sim 10^7$ A/cm$^2$ if $w$ and
$t$ are properly tuned. Finally the report~\cite{Yamanouchi} of
the reduced $J_{\rm c}$ in ferromagnetic semiconductors is yet
limited to low temperatures ($\sim$ 100 K) while the reduction
scheme presented in this Letter does not require low temperature
operation.

In summary, we have clarified the mechanism by which the PMA can
drastically reduce $J_{\rm c}$. The geometrical tuning is
important to maximize the reduction by the PMA. When properly
tuned, the dependence of $J_{\rm c}$ on $\beta$ and the DW pinning
force $f_{\rm pin}$ is very weak. This result solves the large
thermal fluctuation problem and also makes feasible nanoscale
magnetoelectronic devices~\cite{Parkin,Allwood} based on the
current-induced DW motion.

Two authors (SWJ and WK) contributed equally to this work.
Critical comments by Sug-Bong Choe, Jae-Hoon Park, and Tae-Suk Kim
are appreciated. This work was supported by the KOSEF (SRC Program
No. R11-2000-071, Basic Research Program No. R01-2007-000-20281-0,
NRL Program No. M10600000198-06J0000-19810), the KRF (No.
KRF-2006-311-D00102), and POSTECH (Core Research Program). We
acknowledge the support by the KISTI under the Strategic
Supercomputing Support Program with Sik Lee as the technical
supporter. The use of the computing system at the Supercomputing
Center is appreciated.

\newpage

\newpage

\begin{thebibliography}{99}
\bibitem{Jc-minimum-comment} In Figs.~\ref{simulation}(b) and (c),
$J_{\rm c}$ is minimized around $w-\delta w/2$ (or $w_{\rm ave}$)
near 72 nm, which is somewhat smaller than the value ($\sim 78$
nm) in Fig.~\ref{analytical}(b). We attribute this difference to
the pinning-induced DW deformation.
\end{thebibliography}
\end{document}